\documentclass[twocolumn]{aastex631}
\usepackage{natbib}
\usepackage{amsmath}
\usepackage{bm}
\usepackage{graphicx}
\usepackage{multirow}
\usepackage{booktabs} 
\usepackage{array}
\usepackage{xcolor} 
\usepackage{hyperref}

\begin{document}

\title{A Data-constrained Magnetohydrodynamic Simulation of Successive X-class Flares in Solar Active Region 13842 I. Dynamics of the Solar Eruption Associated with the X7.1 Solar Flare}

\author[0000-0003-2002-0247]{Keitarou Matsumoto}
\affiliation{Center for Solar-Terrestrial Research, New Jersey Institute of Technology, University Heights, Newark, NJ 07102-1982, USA}
\author[0000-0001-5121-5122]{Satoshi Inoue}
\affiliation{Center for Solar-Terrestrial Research, New Jersey Institute of Technology, University Heights, Newark, NJ 07102-1982, USA}
\author[0000-0002-6018-3799]{Nian Liu}
\affiliation{Center for Solar-Terrestrial Research, New Jersey Institute of Technology, University Heights, Newark, NJ 07102-1982, USA}
\author[0000-0001-9046-6688]{Keiji Hayashi}
\affiliation{Center for Computational Heliophysics, New Jersey Institute of Technology, University Heights, Newark, NJ 07102-1982, USA}

\author[0000-0002-8179-3625]{Ju Jing}
\affiliation{Center for Solar-Terrestrial Research, New Jersey Institute of Technology, University Heights, Newark, NJ 07102-1982, USA}
\author[0000-0002-5233-565X]{Haimin Wang}
\affiliation{Center for Solar-Terrestrial Research, New Jersey Institute of Technology, University Heights, Newark, NJ 07102-1982, USA}

\email{km876@njit.edu} 



\begin{abstract}
We investigated the initiation and the evolution of an X7.1-class solar flare observed in solar active region NOAA 13842 on October 1, 2024, based on a data-constrained magnetohydrodynamic (MHD) simulation. The nonlinear force-free field (NLFFF) extrapolated from the photospheric magnetic field about 1 hour before the flare was used as the initial condition for the MHD simulations. The NLFFF reproduces highly sheared field lines that undergo tether-cutting reconnection in the MHD simulation, leading to the formation of a highly twisted magnetic flux rope (MFR), which then erupts rapidly driven by both torus instability and magnetic reconnection. This paper focuses on the dynamics of the MFR and its role in eruptions. We find that magnetic reconnection in the pre-eruption phase is crucial in the subsequent eruption driven by the torus instability. Furthermore, our simulation indicates that magnetic reconnection also directly enhances the torus instability. These results suggest that magnetic reconnection is not just a byproduct of the eruption due to reconnecting of post-flare arcade, but also plays a significant role in accelerating the MFR during the eruption. 
\end{abstract}

\keywords{Solar flares (1496) --- Magnetohydrodynamics(1964) --- Solar active region
magnetic fields (1975) --- Magnetohydrodynamical simulations(1966)}


\section{Introduction} \label{sec:intro}
It is widely believed that solar flares are caused by magnetic reconnection (\citealt{Carmichael1964}; \citealt{Sturrock1966}; \citealt{Hirayama1974}; \citealt{Kopp1976}). The magnetic energy released during the reconnection process is primarily converted into thermal and kinetic energy, allowing us to observe flares and coronal mass ejections (CMEs) (\citealt{Fletcher2011}) at various wavelengths. Solar flares associated with CMEs are extremely important in space weather, causing magnetic storms and ground-level enhancements (GLE) (\citealt{Gonzalez1994}; \citealt{Shea2012}). Therefore, from the perspective of space weather, it is crucial to understand the entire process of solar flares from their birth to their evolution. 

Associated with solar flares, the eruption of a helical-structured object is often observed. This helical structure is a bundle of twisted magnetic field lines, called a magnetic flux rope (MFR), which corresponds to the core of the CME (\citealt{Forbes2000}). To explain solar eruptions, many theoretical models have been proposed, some based on ideal magnetohydrodynamic (MHD) instabilities, e.g., the torus instability (TI: \citealt{Bateman1978, Kliem2006}), the kink instability (\citealt{Kruskal1954, Torok2005}), the double-arc instability (\citealt{Ishiguro2017}), and others on magnetic reconnection, e.g., the breakout reconnection (\citealt{Antiochos1999}), the tether-cutting reconnection (\citealt{Moore2001, Jiang2021}). Several models, including both ideal MHD instabilities and reconnection, have also been proposed (\citealt{Aulanier2010,Amari2018,Inoue2018}). However, we have not reached a common understanding.

One of the reasons for this difficulty is the limited observational data, that is, solar observation data cannot provide three-dimensional (3D) information, especially 3D magnetic field information. 
Since magnetic pressure is dominant over plasma pressure in the corona (\citealt{Gary2001}), i.e., it is in an approximate zero-$\beta$ condition, the coronal magnetic field can be described as a ``force-free'' field. 
Therefore, a nonlinear force-free field (NLFFF) extrapolation (\citealt{wiegelmann2012}), which is an extrapolation method of the coronal magnetic field under a force-free approximation, has been a very useful tool for understanding 3D magnetic fields. However, the NLFFF model is a static model and is not suitable for showing the dynamics of the magnetic field during the flares. To overcome this issue, data-constrained simulations have been performed to understand the initiation and dynamics of the coronal magnetic field in solar flares. (\citealt{Jiang2016, Muhamad2017, Inoue2021, Yamasaki2022}).

The solar active region (AR) NOAA 13842 produced two X-class flares (X7.1 and X9.0) in October 2024. Rich data for the photospheric magnetic field from Helioseismic and Magnetic Imager (HMI: \citealt{Scherrer2012}) onboard Solar Dynamics Observatory (SDO: \cite{Pesnell2012}) showed the temporal evolution of magnetic field in high spatial and temporal resolutions. The extreme ultra-violet image from Atmospheric Imaging Assembly (AIA: \citealt{Lemen2012}) onboard the SDO showed the characteristic structure and dynamics in pre- and post-eruption. However, they don't provide quantitative physical values in 3D space. Therefore, we conducted a data-constrained MHD simulation to understand 3D dynamics of the X7.1 flare that occurred on 2024 October 1. We used the NLFFF as an initial condition of the MHD simulation. The rest of this paper is constructed as follows: Observations and simulation methods, results, and discussion are described in Sections \ref{sec:sec2}, \ref{sec:sec3}, and \ref{sec:sec4}, respectively. Finally, the conclusions are summarized in Section \ref{sec:sec5}.

\section{Observations and MHD simulations}
\label{sec:sec2}
\subsection{Observations} 
\begin{figure*}[ht!]
\plotone{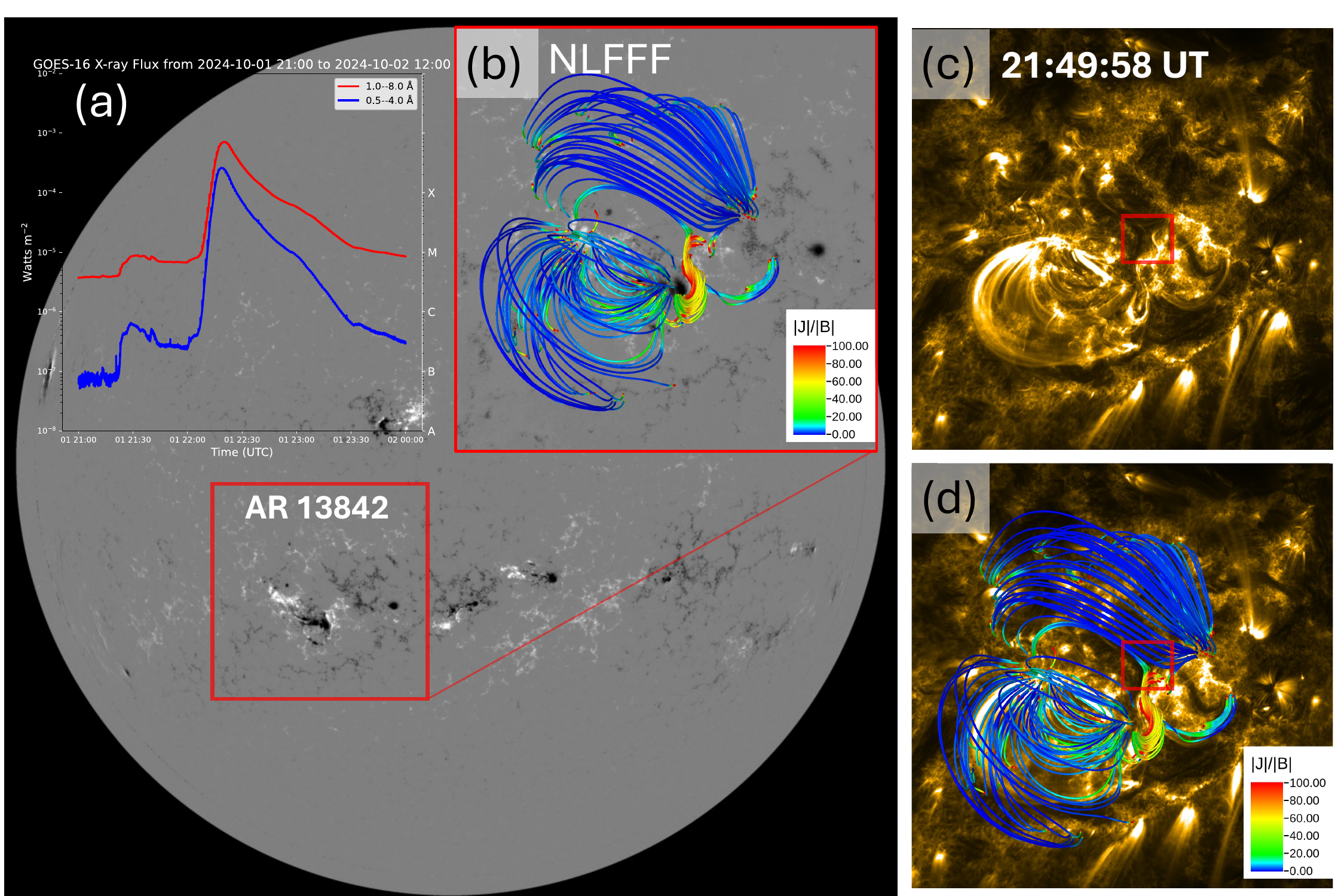}
\caption{Overview of the GOES X-ray flux, the NLFFF, and the AIA 171 \AA\ observations. (a) Time evolution of the X-ray flux measured by the GOES-16 satellite between 21:00 UT on 2024 October 1 and 00:00 UT on October 2. The solar X-ray emissions in the 1.0–8.0 \AA\ and 0.5–4.0 \AA\ passbands are shown in red and blue, respectively. (b) An NLFFF extrapolated from the HMI data at 20:36 UT on 2024 October 1. The background corresponds to the HMI magnetogram at 20:36 UT. (c) AIA 171 \AA\ image observed at 21:49:58 UT on 2024 October 1. (d) Overlay of the NLFFF extrapolated in (b) with the AIA 171 \AA\ image in (c). The region enclosed by the red rectangle in panels (c) and (d) indicates the separatrix structure.
\label{fig:fig1}}
\end{figure*}

The X7.1 flare occurred in NOAA AR 13842 on October 1, 2024. According to the Geostationary Operational Environmental Satellite (GOES) X-ray, the start time and peak time were 21:58 UT and 22:20 UT, respectively, as shown in Figure \ref{fig:fig1} (a). The CME was observed at 23:12 UT  by the Solar and Heliospheric Observatory (SOHO)/Large Angle and Spectrometric Coronagraph Experiment (LASCO) C2 telescope (\citealt{Brueckner1995}). 

To extrapolate the NLFFF, we used the SDO/HMI vector magnetogram, taken at 20:36 UT (about one hour before the flare) and on the cylindrical equal area (CEA) projection, as the bottom boundary condition. The CEA map was generated in the same way as the standard HMI SHARP CEA data series but with an expanded field of view. Figure \ref{fig:fig1} (b) shows the NLFFF extrapolated from HMI data at 20:36 UT. A sigmoidal structure, which refers to an S-shaped or inverse-S-shaped magnetic configuration often associated with twisted magnetic fields and pre-eruption energy storage, is visible along the polarity inversion line (PIL) in this panel. Figure \ref{fig:fig1} (c) displays the AIA 171 \AA\ image at 21:49:58 UT, where similar features of the sigmoid and the separatrix structure, which represents the boundary between different magnetic field lines domains where magnetic reconnection more likely occur, are shown in the red rectangle. In Figure \ref{fig:fig1} (d), the NLFFF in Figure \ref{fig:fig1} (b) is superimposed on the AIA 171 \AA\ image of Figure \ref{fig:fig1} (c) to provide a direct comparison between the modeled magnetic field and the observed coronal structure. This overlay accurately highlights the NLFFF's ability to reproduce the pre-flare magnetic structure, including the sigmoid and separatrix structures.

\subsection{Nonlinear Force-free Extrapolation} 
\label{sec:sec2.2}
To conduct the NLFFF extrapolation and data-constrained MHD simulation, we used the following equations (\citealt{Inoue2016}),
\begin{equation}
\mathit{\rho} = \lvert \bm{B} \rvert,
\label{eq:eq1}
\end{equation}
\begin{equation}
\frac{\partial \bm{v}}{\partial t} = -(\bm{v} \cdot \nabla)\bm{v} + \frac{1}{\mathit{\rho}} \bm{J} \times \bm{B} + \mathit{\nu_{i}} \nabla^2 \bm{v},
\label{eq:eq2}
\end{equation}
\begin{equation}
\frac{\partial \bm{B}}{\partial t} = \nabla \times (\bm{v} \times \bm{B})+\eta_i \bm{\nabla}^2\bm{B} - \nabla \bm{\phi},
\label{eq:eq3}
\end{equation}
\begin{equation}
\bm{J} = \nabla \times \bm{B},
\label{eq:eq4}
\end{equation}
\begin{equation}
\frac{\partial \bm{\phi}}{\partial t} + c_h^2 \nabla \cdot \bm{B} = - \frac{c_h^2}{c_p^2} \bm{\phi},
\label{eq:eq5}
\end{equation}

where $\mathit{\rho}$, $\bm{B}$, $\bm{v}$, $\bm{J}$, and $\mathit{\phi}$ are the plasma density, the magnetic flux density, the velocity, the electric current density, and the scalar potential, respectively. The length, the magnetic field, the plasma density, the velocity, the time, and the electric current density are normalized by $L^{*}$ = 362.5 Mm, $B^{*}$ = $2.915\times 10^{-1}$ T, $\rho^*$ (kg m$^{-3})$ which is the density at the bottom surface in the simulation box, $V_A^* = {B^*}/{(\mu_0 \rho^*)^{1/2}}$, where $\mu_0$ is the magnetic permeability, $\tau_A^* = {L^*}/{V_A^*}$ (s), and $J^* = {B^*}/{\mu_0 L^*}$ (A m$^{-2}$), respectively.
Plasma density, $\mathit{\rho}$, is proportional to $\lvert \bm{B} \rvert$, which was introduced as Alfvén wave propagates faster to weaker magnetic field region. The scalar potential $\phi$ is brought conveniently to reduce the error of ${\bf \nabla }\cdot{\bm{B}}$ (\citealt{Dedner2002}). The coefficients $\nu_i$ and $\eta_i$ are viscosity and electric resistivity where the index $i$ corresponds to NLFFF or MHD. In the NLFFF extrapolation, $\nu_{NLFFF}=1.0\times 10^{-3}$ and $\eta_{NLFFF}=5.0\times 10^{-5}+1.0\times 10^{-3}|{\bm{J}}\times {\bm{B}}||{\bm{v}}|^2/|{\bm{B}}|^2$. The second term of $\eta_{NLFFF}$ is added to speed up the process of reaching the force-free state.
The coefficients ${c_h^2}$ and ${c_p^2}$ in Equation (\ref{eq:eq5}) are fixed at the constant values of 0.04 and 0.1, respectively.

The initial condition was given as the potential field extrapolated from observed $B_{z}$ using the Green function method (\citealt{Sakurai1982}). Regarding the boundary condition, the normal components of the magnetic fields are fixed at all boundaries, while the tangential components follow the induction equation, except at the bottom boundary. The velocities are fixed to zero at all boundaries while $\partial / \partial n$ = 0 is applied to $\phi$. Specifically, the tangential component of the magnetic field at the bottom boundary is according  to the following equation, 
\begin{equation}
\bm{B}_{\text{bc}} = \gamma \bm{B}_{\text{obs}} + (1 - \gamma) \bm{B}_{\text{pot}},
\end{equation}
where $\bm{B}_{\text{bc}}$ corresponds to the tangential component, which represents a linear combination of the observed magnetic field ($\bm{B}_{\text{obs}}$) and the potential magnetic field ($\bm{B}_{\text{pot}}$). $\gamma$ is a parameter ranging from 0 to 1. Initially, the parameter $\gamma$ is set with zero and updated as  $\gamma + d\gamma$ during the iteration when the total Lorentz force, $R = \int |\bm{J} \times \bm{B}|^2 dV$, falls below a critical threshold defined with $R_{\text{min}}$. The magnetic field is fixed with the observed one after $\gamma$ could reach 1. In this study, we used $R_{{\text{min}}}=5.0 \times 10^{-3}$ and $d\gamma =0.02$. Additionally, we limit the velocity to avoid sharp discontinuities, especially between the boundary and inner region. If Alfvén Mach number $v^* \left(= {|\bm{v}|}/{|\bm{v}_{\text{A}}|} \right)$ exceeds a specified limit $v_{\text{max}}$ (set to 0.04 in our case), we adjust the velocity using the relation $\bm{v} \rightarrow \left( v_{\text{max}}/v^* \right) \bm{v}$. This approach prevents abrupt changes in the velocity from propagating into the domain during the iterative process.

  \subsection{Data-constrained MHD Simulation} 
Next, we conducted data-constrained MHD simulations using the NLFFF as the initial condition to trace the dynamics from the initiation to the eruption. Although the equations are the same as one for the NLFFF, the primary difference is the bottom boundary condition for the tangential components of the magnetic field. The normal component of the magnetic field at the bottom boundary is fixed in time, while the tangential components are allowed to evolve freely, following the induction equation where all the velocity components are fixed with zero. In the MHD simulation, the resistivity  $\eta_{\text{MHD}}$ and the viscosity $\nu_{\text{MHD}}$ were set to constant values of $1.0 \times 10^{-5}$ and $1.0 \times 10^{-4}$, respectively. The MHD simulations do not limit the magnitude of $\bm{v}$. 

The numerical simulation box of both NLFFF and MHD simulation was assigned with 362.5 $\times$ 362.5 $\times$ 362.5 $\mathrm{Mm}^3$ which corresponds to 1.0 $\times$ 1.0 $\times$ 1.0 in the non-dimensional unit. Although original HMI data was assigned with $1000 \times  1000$ gird points, the $2 \times 2$ binning process proceeded, resulting in $500 \times 500$ being assigned. The simulation time $t$ in this study is normalized by the Alfvén time (see Section \ref{sec:sec2.2}), where $t = 1$ corresponds to approximately 6 minutes in physical time.

  \section{Results}
  \label{sec:sec3}
\subsection{Physical condition of the Initial MFR}\label{sec:sec3.1}

\begin{figure*}[ht!]
\plotone{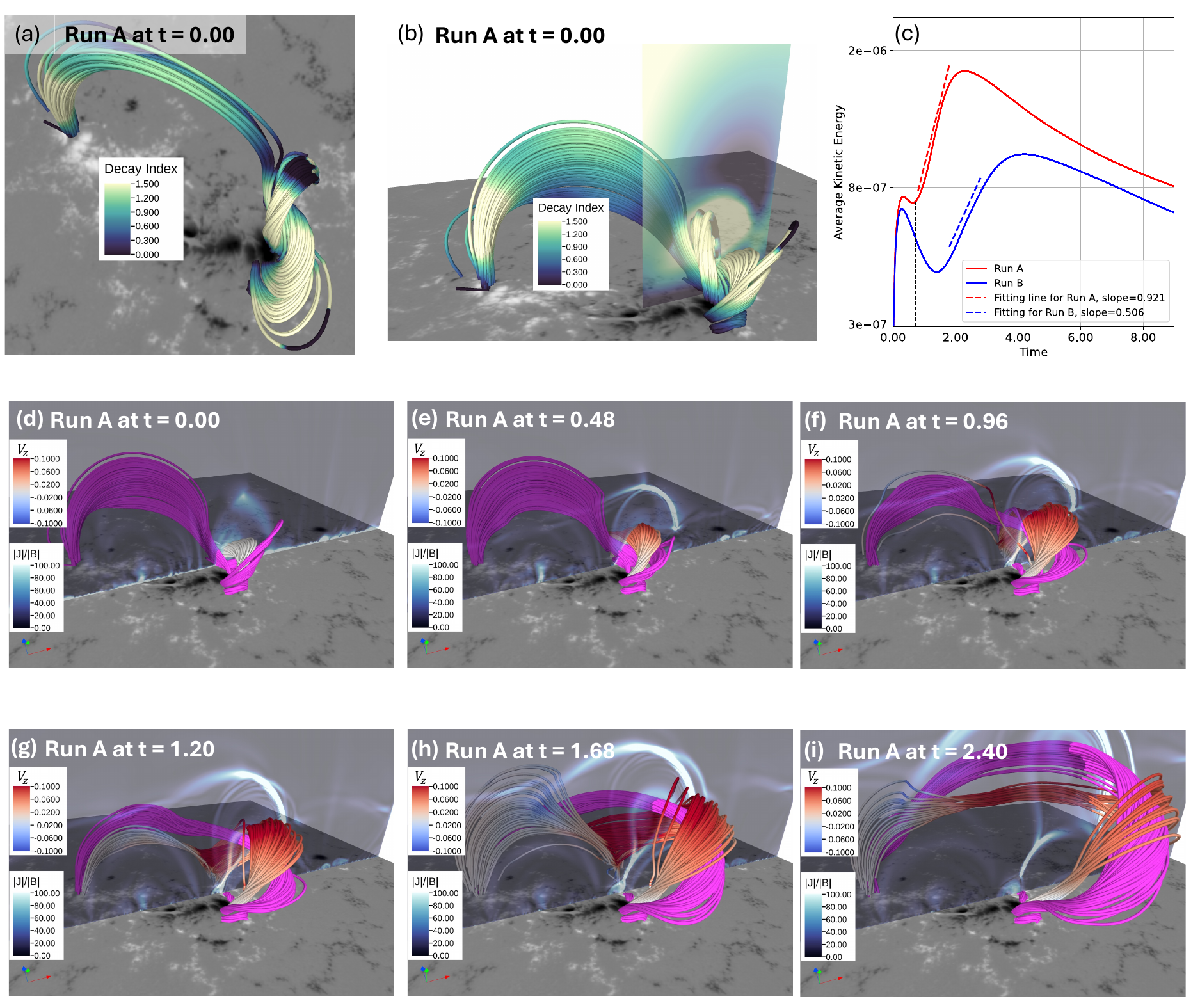}
\caption{(a) Twisted field lines with $T_{W}$ exceeding 0.5 in the NLFFF, which correspond to \(t = 0\) in Run A. The color of the field lines represents the decay index. (b) Side view of the panel (a) except that the decay index is plotted on the vertical cross-section. (c) Temporal evolution of the averaged kinetic energy for Runs A and B, respectively. Black dashed lines in Runs A  and B indicated the time at \(t = 0.72\) and at \(t = 1.44\), respectively. These times show the end of the pre-eruption phase and the start of the erupting phase discussed in Section \ref{sec:sec4}. (d)--(i) Temporal evolution of the twisted field lines for Run A. The vertical cross-section represents \(|\bm{J}|/|\bm{B}|\). Purple twisted field lines are part of the sheared field lines in the NLFFF and demonstrate the tether-cutting reconnection. They ultimately ascend as the MFRs. The field lines colored by \(V_z\) satisfy the decay index of 1.5 or greater at $t=0$ and potentially destabilize to TI. (An animation of this figure is available)
\label{fig:fig2}}
\end{figure*}

We calculate the twist of the field lines to address the non-potentiality of the NLFFF and discuss the stability, which was used as the initial condition of the MHD simulation. The magnetic twist, $T_w$, is defined as the following,
\begin{equation}T_w = \frac{1}{4\pi} \int \frac{{\bm J}\cdot {\bm B}}{|\bm{B}|^2} \, dl,
\label{eq:eq8}
\end{equation}
where $dl$ is a line element (\citealt{Berger2006}) and $T_w$ is calculated for each field line. The magnetic twist has positive and negative signs, which represent the sign of the magnetic helicity. Note that $T_w$ measures the number of turns of two infinitesimally close field lines, which is distinct from the number of turns of the field lines around the magnetic axis of the MFR. Figure \ref{fig:fig2} (a) shows the selected field lines in the NLFFF that satisfy more than $T_w=0.5$. Since the sigmoidal structure formed in Figure \ref{fig:fig2} (a) shows an S-shaped structure, this indicates that positive helicity is dominant over the active region. According to the NLFFF analysis done in \cite{Inoue2011, Inoue2013}, the twisted field lines with more than $|T_w|=0.5$ mostly disappeared after the flare. They suggest that the twisted field lines with more than $|T_w|=0.5$ were involved in flare occurrences. Therefore, this study focused on the field lines, which satisfy $T_w \ge 0.5$, in the pre-flare phase to trace the dynamics in the MHD simulation.

Next, we calculated the decay index to discuss the stability of the TI (\citealt{Bateman1978, Kliem2006}). 
The decay index, $n$, is expressed by the following equation, 
\begin{equation}
n = -\frac{d \ln \bm{B}_{\text{ex}}}{d \ln z},
\label{eq:eq9}
\end{equation}
where $\bm{B}_{\text{ex}}$ is the horizontal components of the external field. The decay index represents how rapidly the horizontal component of the external magnetic field decays in the height direction. In this study, the external magnetic field approximates the potential field. The threshold of the TI is well known as $n \approx1.5$. When the axis of the MFR exceeds the height that satisfies $n \approx 1.5$, the hoop force working on the MFR is dominant over the suppressing force coming from the external field, resulting in an eruption of the MFR. Figure \ref{fig:fig2} (a) shows that the top of the twisted field lines with more than $T_w > 0.5$ have nearly reached a critical height of the TI. However, since the decay index that determines the instability will strongly depend on the boundary condition and the structure of the MFR, this result only suggests that these field lines may potentially be destabilizing (\citealt{Olmedo2010, Zuccarello2015, Alt2021}). Furthermore, we found that a stable region to the TI (where $n < 1.5$) exists above the twisted field lines in Figure \ref{fig:fig2} (b). Therefore, an MHD simulation that can clear the temporal evolution of the magnetic field is needed to clarify the stability.

\subsection{Overview of 3D Dynamics of the Magnetic Field Lines}
\label{sec:sec3.2}
To investigate the initiation and evolution of the twisted field lines, we performed a data-constrained MHD simulation using the NLFFF as the initial condition. This simulation is referred to as Run A. Figure \ref{fig:fig2} (c) shows the temporal evolution of the kinetic energy ($\int (\rho |\bm{v}|^2 / 2) \, dV$, where $dV$ is a volume element) for Runs A and B, the latter of which will discuss later. Figures \ref{fig:fig2} (d)-(i) show the temporal evolution of the magnetic fields in Run A. The vertical cross-section represents \(|{\bm{J}}|/|{\bm{B}}|\) to enhance the current sheet. We plotted the field lines in Figures \ref{fig:fig2} (d)-(i), selected from the twisted field lines shown in Figure \ref{fig:fig2} (a) for easier viewing. The purple twisted field lines underwent the tether-cutting reconnection in Figures \ref{fig:fig2} (d)-(f) (\citealt{Moore1980}; \citealt{Moore2001}) above the PIL, resulting in more highly twisted field lines and ultimately erupting as the MFR in Figures \ref{fig:fig2} (g)-(i). We plot the colored twisted field lines above the purple field lines with $V_z$. As discussed in Section \ref{sec:sec3.1}, these field lines were suggested to be either destabilized or close to the unstable state. Whichever state these field lines are in, the newly formed MFR via tether-cutting reconnection destabilizes them by pushing them up. Eventually, the colored field lines reconnected with other field lines, contributing to the twist in the erupting MFR and forming a stronger MFR. 
\begin{figure*}[ht!]
\plotone{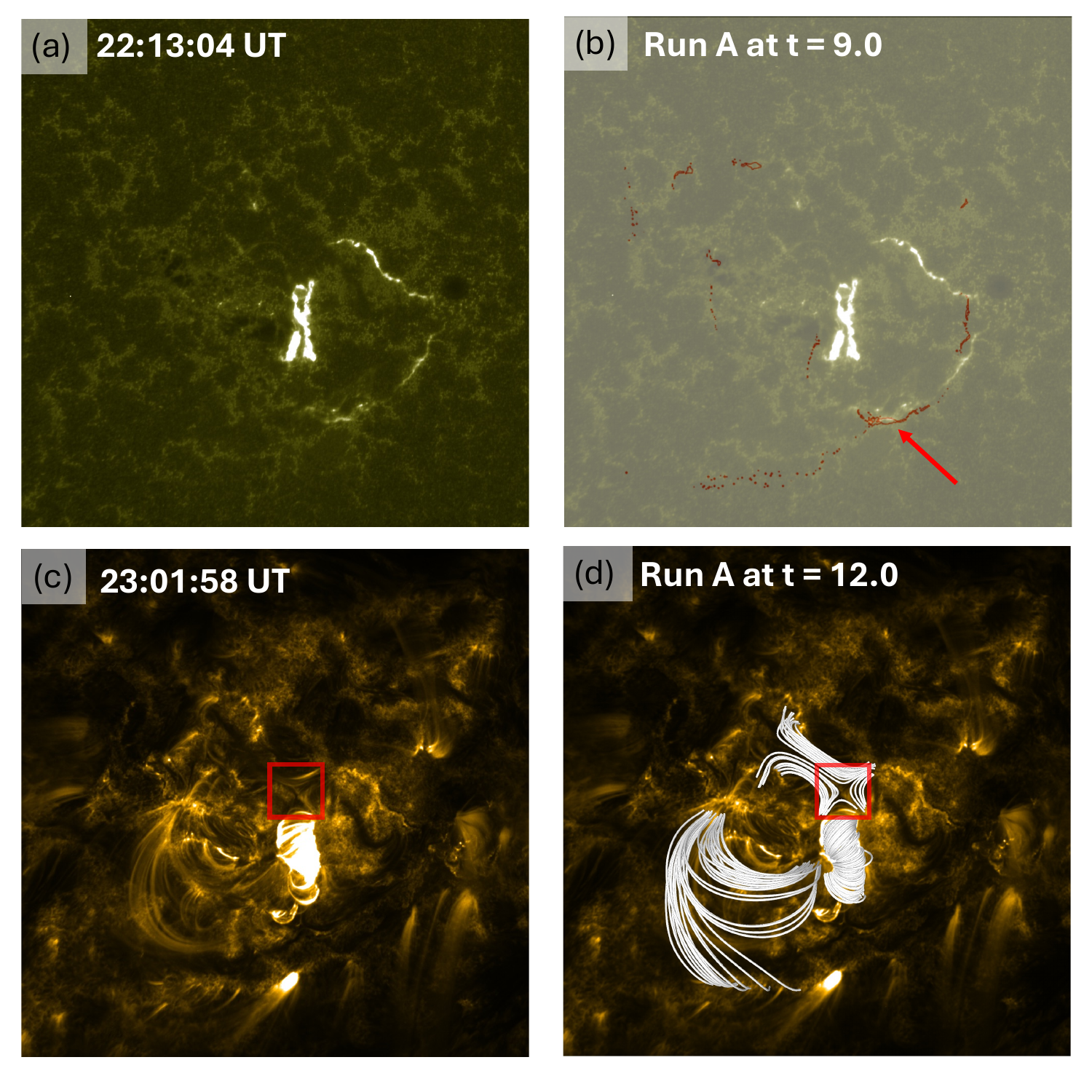}
\caption{(a) AIA 1600\,\AA\ image at 22:13:04 UT. (b) Contours of \(T_w = 0.85\) at \(t = 9.0\), overlaid on (a). The red arrows show the region corresponding closely to the location of the remote brightening. (c) AIA 171\,\AA\ image at 23:01:58 UT. (d) The field lines at \(t = 12.0\) are superimposed on (c). The region enclosed by the red square in panels (c) and (d) indicates the separatrix structure suggested by the AIA image. (An animation of AIA 1600\,\AA\ is available)
\label{fig:fig3}}
\end{figure*}

\subsection{Comparison with the Observation (AIA)}\label{sec:sec3.3}
To validate our simulation, we compared our simulation results with the AIA 1600 \AA\ and 171 \AA. Figure \ref{fig:fig3} (a) shows the AIA 1600 \AA\ image at 22:13:04 UT, and Figure \ref{fig:fig3} (b) shows that the twist distribution at $t = 9.0$, which corresponds to the footpoints of the field lines with $T_w > 0.85$, superimposed on the AIA image. We selected $T_w > 0.85$ to enhance the footpoints of erupting MFR. Figure \ref{fig:fig3} (a) shows the typical two-ribbon flares often observed after flares. Additionally, remote brightening is visible a short distance from the flare ribbons. In Figure \ref{fig:fig3} (b), the locations of the twist distributions coincide well with part of remote brightening. This result indicates that part of remote brightening locations can cover the footpoints of the erupting MFR. Figure \ref{fig:fig3} (c) shows the AIA 171\AA\ image at 23:01:58 UT, and Figure \ref{fig:fig3} (d) shows the magnetic field lines at \(t=12.0\), superimposed on Figure \ref{fig:fig3} (c). Our simulation reproduced post-flare loops shown in Figure \ref{fig:fig3} (c), where the footpoints of the post-flare loops coincided well with the location of two-ribbon flares in Figure \ref{fig:fig3} (a). These results indicate that the bright regions observed at 1600 \AA\ following the flare correspond to the footpoints of post-flare loops observed as two-ribbon flares and to that of the erupting MFR observed as part of remote brightening. Additionally, the field line structure shown in the red square in Figure \ref{fig:fig3} (d) shows a separatrix structure corresponding to the observations suggested by the same square in Figure \ref{fig:fig3} (c). 
In Figures 2 (d)-(i), the tether-cutting reconnection occurred at the strong current sheet, and the footpoints of the MFR were anchored on the active region and its northeast location. From the discussion above, some footpoints of the MFR should be anchored on the remote ribbons in the west of the active region. The separatrix structure has a null point, which may help the footpoints shift to the western region.

\section{Discussion}
\label{sec:sec4}
\subsection{A Role of the Reconnection in Pre-eruption Phase}
\label{sec:sec4.1}
\begin{figure*}[ht!]
\plotone{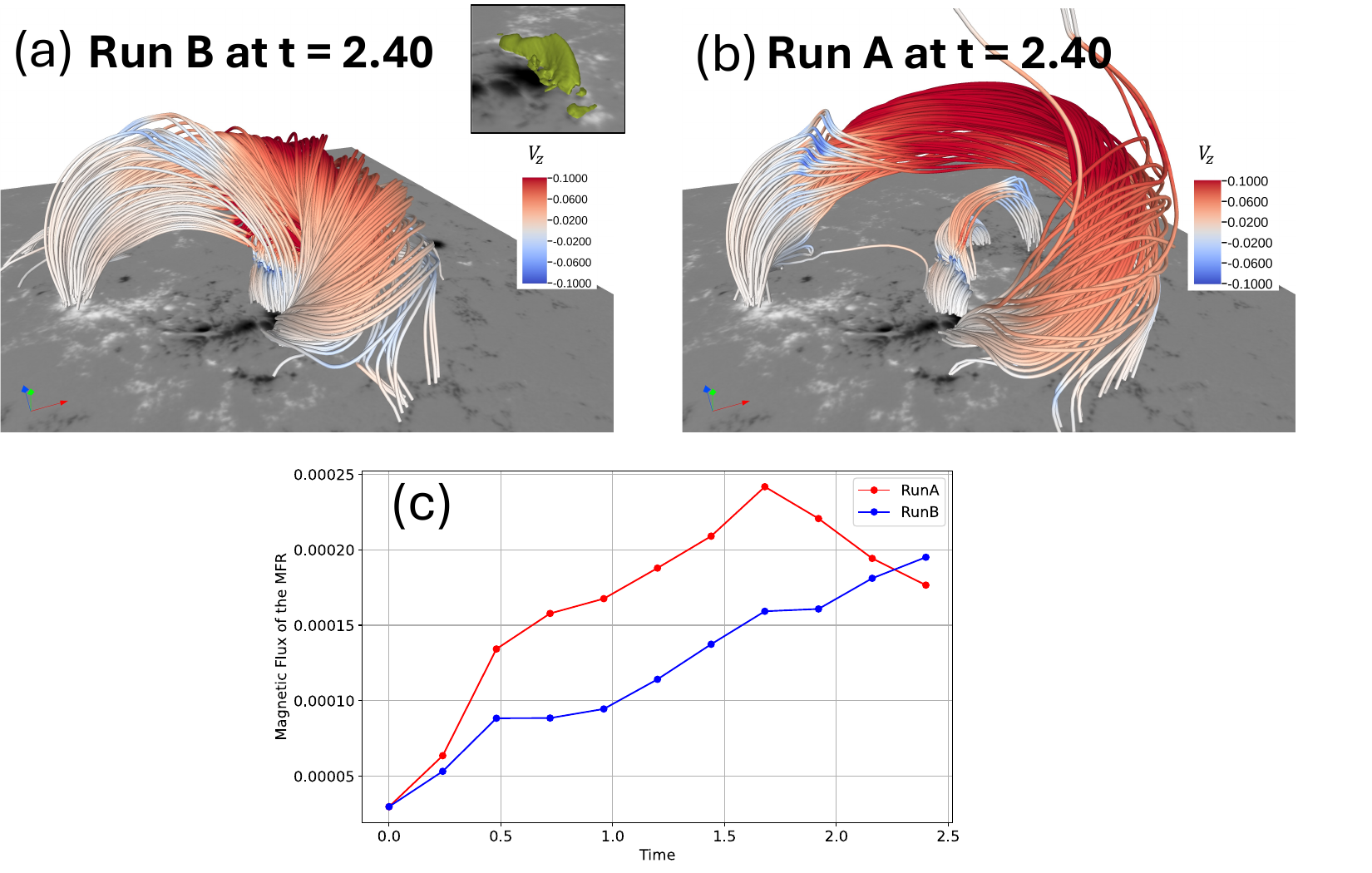}
\caption{Panels (a) and (b) show the twisted field lines whose color represents $V_z$ in Runs B and A, respectively. (a) The magnetic field obtained from Run B at $t=2.40$. The small insertion in the top-right shows the iso-surface of $|\bm{J}|=30$ in the NLFFF ($t=0.0$). (b) The magnetic field obtained from Run A at $t=2.40$. (c) Temporal evolution of the magnetic flux of the MFR, which focused on the twisted field lines with $T_w > 1.25$, in Runs A and B, respectively. (An animation of this figure is available)
\label{fig:fig4}}
\end{figure*}

From the results of Section \ref{sec:sec3.2}, there were two key components for the eruption in our simulation: one is the tether-cutting reconnection, and the other is the torus instability. However, it is unclear how exactly these are involved during the eruption. To clear this, we ran another simulation, Run B, where the magnetic reconnection was suppressed by forcing the velocity to stop at strong current density (\citealt{Inoue2006, Inoue2018, Yamasaki2022}). Magnetic reconnection typically occurs in strong current sheets, facilitated by plasma inflows and outflows. We forced the velocity to set zero in the high current density region. This process suppresses the plasma motion necessary for reconnection, effectively preventing it. Reconnection was suppressed at \(|\bm{J}| > 30\), which is plotted as iso-surface shown in a small insertion in Figure \ref{fig:fig4} (a). Note that the reconnection was suppressed at the specified current density region, and not all reconnection was suppressed.

Figures \ref{fig:fig4} (a) and (b) show the magnetic structures at $t=2.40$ for Runs A and B, respectively. Although the dynamics of the eruption are the same (see animation of Figure \ref{fig:fig4}), the velocity of the twisted field lines in Run B is slower due to the suppressed reconnection. Figure \ref{fig:fig4} (c) shows the temporal evolution of the magnetic flux in the highly twisted field lines. This flux follows the equation, \(\Phi = \int_{T_w > 1.25} B_z \, dS\), where \(B_z\) corresponds to the normal component of the positive magnetic field at the bottom boundary. The reason for selecting \(T_w > 1.25 \) is that the magnetic flux occupied with the field lines with $T_w > 1.25$ increased by $t=2.7$. Meanwhile, the magnetic flux occupied with the field lines with $0.5 < T_w < 1.25$ decreased. This means that some field lines with twist $0.5 < T_w  < 1.25$ were converted to erupting field lines with twist $T_w > 1.25$ through the reconnection. Therefore, we used \(T_w > 1.25\) to calculate \(\Phi\). Obviously, the growth of the MFR was suppressed in Run B, as shown in Figure \ref{fig:fig4} (c). Figure \ref{fig:fig2} (c) plots the kinetic energy, indicating that Run B shows a lower energy level during the pre-eruption phase, which is mentioned in the caption of Figure \ref{fig:fig2} (c), due to suppressed reconnection. Furthermore, this suppression affects the growth of the instability. The kinetic energy is presented on a logarithmic scale, and the growth rate of the instability is represented by linear dashed slopes in red and blue. Since the dynamics are already in a nonlinear phase, this growth rate differs from that obtained through linear stability analysis. Nevertheless, it is a valuable quantitative measure for comparing Runs A and B. 

This study's important finding is that the tether-cutting reconnection in the pre-eruption phase is critical to the subsequent eruption driven by the TI. If the reconnection occurs between the sheared field lines, the twisted field lines are created, which increases the toroidal current within the MFR, resulting in the enhancement of the hoop force acting on it. Therefore, the reconnection process in the pre-eruption phase is essential to determine the subsequent eruption driven by the TI. 

\subsection{A Role of Reconnection in Erupting Phase}
\label{sec:sec4.2}
\begin{figure*}[ht!]
\plotone{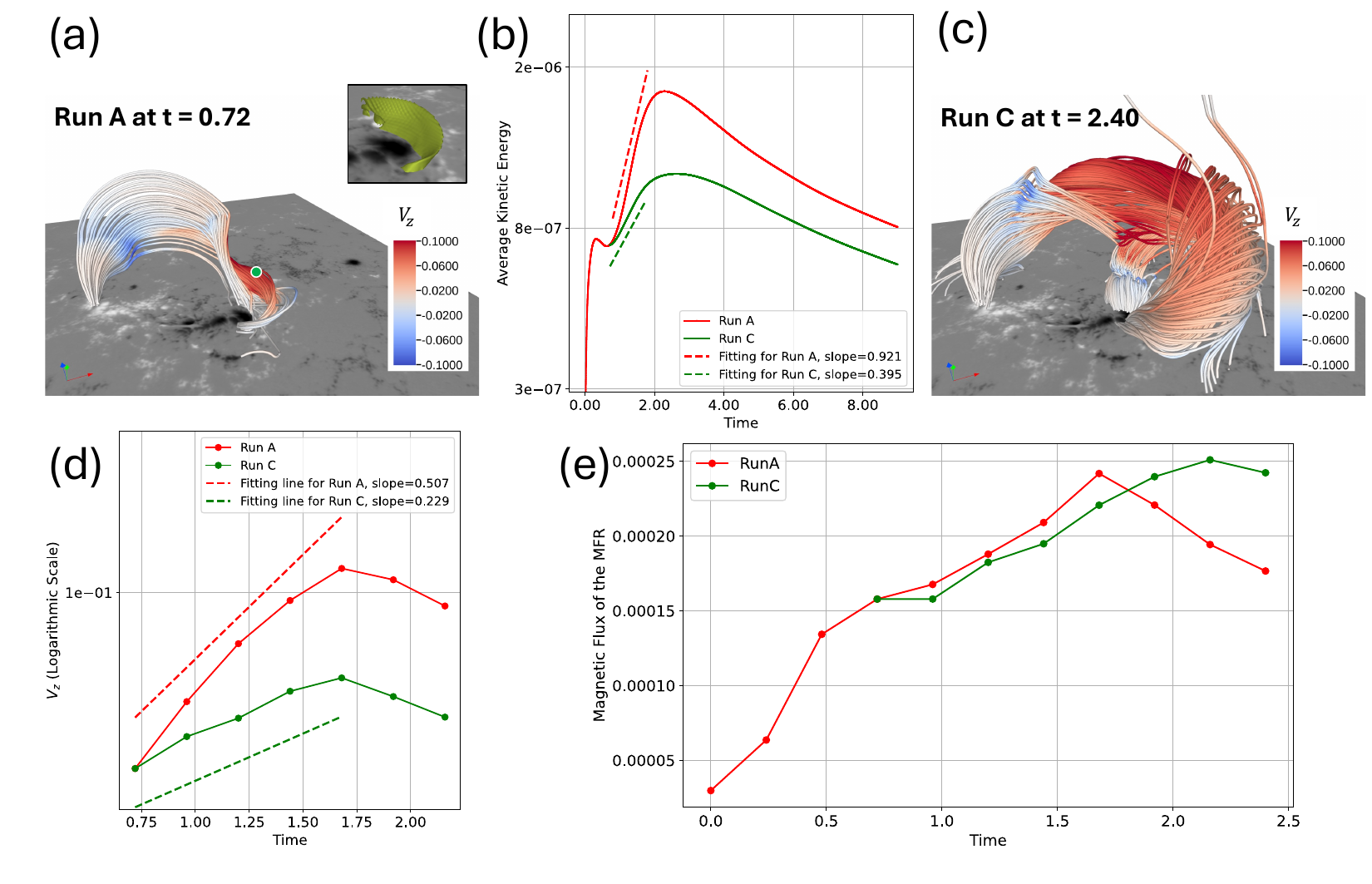}
\caption{(a) Twisted field lines in Run A at \(t = 0.72\). The small insertion in the top-right shows the iso-surface of $|\bm{J}|=25$ in Run A at $t=0.72$, where the reconnection is suppressed in Run C. The green circle indicates the location \((x, y, z) = (0.572, 0.418, 0.680)\), where we put a test particle at \(t = 0.72\) to trace the evolution of the twisted field lines using Lagrangian tracking. (b) Temporal evolution of the averaged kinetic energy for Runs A and C, respectively. (c) Twisted field lines in Run C at \(t = 2.40\). (d) Temporal evolution of the \(V_z\) of the MFR as a result of Lagrangian tracking. (e) Temporal evolution of the magnetic flux of the MFR in Runs A and C, respectively, where the twisted field lines with $T_w > 1.25$ are focused. (An animation of this figure is available)
\label{fig:fig5}}
\end{figure*}
We investigate the role of the reconnection in the MFR's eruption driven by the TI in erupting phase of Run A. We performed a new simulation, Run C, where the reconnection was suppressed after \(t = 0.72\) in Run A using a similar method. Figure \ref{fig:fig5} (a) shows the magnetic twisted field lines at $t=0.72$ and the magnetic reconnection that occurs under them is suppressed. Note that the reconnection was suppressed at \(|\bm{J}| > 25\), which is shown in a small insertion of Figure \ref{fig:fig5} (a). Figure \ref{fig:fig5} (b) shows the kinetic energy and the growth of the instability, evaluated by the slope indicated by the linear lines, was found to be about $2.3$ times greater for Run A than for Run C. Figures \ref{fig:fig4} (b) and \ref{fig:fig5} (c) show magnetic field structures at $t=2.40$ for Runs A and C, respectively. The velocity on top of the MFR appears to be more enhanced in Run A. Therefore, we found that reconnection in the erupting phase plays a role in accelerating the MFR driven by the TI (\citealt{Welsch2018, Inoue2018}).

However, the kinetic energy was calculated in an entire numerical box that includes velocities other than the velocity of the MFR. To assess the MFR velocity more accurately, we traced a plasma element on the MFR using Lagrangian tracking. The Lagrangian tracking is based on the velocity definition \(\bm{v} = \frac{d\bm{r}}{dt}\), where $\bm{r}$ is the position. The position of the plasma element was updated using \({\bm{r}} = \bm{r}_{0} + \bm{v}\,dt\). At \(t = 0.72\), the plasma element was initially placed at \((x, y, z) = (0.572, 0.418, 0.680)\) in Runs A and C, as shown in Figure \ref{fig:fig5} (a). Since coronal plasma is frozen into the magnetic field lines, we can consider that the velocity of the plasma element corresponds to the local velocity of the MFR. Figure \ref{fig:fig5} (d) shows the temporal evolution of the local velocity of the MFR for Runs A and C, respectively. The growth rates of the instability in Runs A and C differ by about $2.2$, which is consistent with what is measured from the kinetic energy. Figure \ref{fig:fig5} (e) shows the temporal evolution of the magnetic flux of the MFR. In Run C, the reconnection appears to be slightly suppressed despite the marked difference in both kinetic energy profiles. As the MFR rises, a vertical current sheet forms beneath them. However, if the magnitude of the current density does not reach the threshold of \( |\bm{J}| = 25 \), the reconnection is allowed. The magnitude of the current density weakened as it extended vertically upward, and then the reconnection occurred at an upper region of the current sheet formed in the weak magnetic fields in Run C. This suggests that suppressing the reconnection in the early phase is crucial for the eruption. In other words, the reconnection involved in the strong magnetic fields is essential for the acceleration.

\section{Summary}
\label{sec:sec5}

We investigated the initiation and the eruption of the X7.1-class solar flare observed on 2024~October~1 in NOAA AR~13842. We successfully reproduced observed phenomena during the X7.1 flare using data-constrained MHD simulations. The X7.1 flare produced typical two-ribbon flares and remote brightening regions at a distance from the PIL. We found that part of the remote brightening regions observed at 1600\,\AA\ correspond to the locations where the footpoints of the MFR are anchored, while the footpoints of the post-flare loops are anchored on the flare ribbons. In addition, we reproduced a separatrix structure on the north area of the flare ribbons. Our simulation focused on the tether-cutting reconnection and the formation of the MFR, whose footpoints are anchored in the northeast area and the active region. On the other hand, we found that some footpoints are anchored on the remote brightening region. This would be due to the reconnection at the null point in the separatrix structure.
 Our simulation indicated that the tether-cutting reconnection in the pre-eruption stage is crucial in accelerating the MFRs. Since the initiation of this flare is driven by the tether-cutting reconnection, the supplied twisted field line by the reconnection enhances the toroidal current flowing inside the MFRs, resulting in enhancing upward hoop force. Therefore, understanding how electric current accumulates in the MFRs during the pre-eruption phase is crucial for the subsequent major eruption. Furthermore, we found that reconnection is essential, even during the erupting phase. The findings suggest that magnetic reconnection significantly contributes to the acceleration of the MFR, rather than being a secondary consequence of the eruption.

In this study, the reconnection occurred automatically at a strong current density region because the resistivity $\eta$ in the induction equation worked efficiently. Therefore, unfortunately, we cannot pinpoint the exact location of the triggering process of the X7.1 flare. Nevertheless,  the importance of reconnection in the scheme of tether-cutting is demonstrated.  In future work, we will focus on the flare trigger of this X7.1 as well as the X9.0 flares in Oct 2024 by analyzing the data from \textit{Hinode}/Solar Optical Telescope (\citealt{Tsuneta2008}) and SDO/HMI. We will comprehensively understand both flares through numerical simulation and data analysis, from the triggering process to the large-scale eruption.

\begin{acknowledgments}

This study is supported by NASA grants 80NSSC23K0406, 80NSSC21K1671, 80NSSC21K0003, 80NSCC24M0174, and NSF grants AST-2204384,  2145253, 2149748, 2206424, 2309939 and 2401229
  The 3D visualizations were produced using VAPOR (\href{http://www.vapor.ucar.edu}{\texttt{www.vapor.ucar.edu}}), a product of the National Center for Atmospheric Research (\citealt{Li2019}). 
\end{acknowledgments}

\bibliography{sample631}{}
\bibliographystyle{aasjournal}



\end{document}